\documentclass{article}
\usepackage{amssymb,amsmath,amsfonts}

\begin{document}

\title{Long-range Correlations in a Data Sequence Extracted Via Polymeric Compaction}
\author{Theo Odijk\\
Lorentz Insitute for Theoretical Physics\\
University of Leiden\\
The Netherlands\\
E-mail: sageron09@gmail.com}

\maketitle

\begin{abstract}
A numerical method is proposed to remove the quenched randomness from a data
sequence of numbers. A polymer chain of beads is introduced with both a hard
core interaction and an appropriate energy associated with the data sequence.
The quenched randomness is hypothesized to collapse the chain to a spherical
globule. Long-range informational correlations then show up in deviations from 
the spherical shape. The resulting coefficients within an expansion in terms of
spherical harmonics go beyond the usual concept of algorithmic information
or Kolmogorov complexity. 
\end{abstract}

\noindent
The Kolmogorov complexity $K$ of a sequence $S_m \!=\! (010110 \ldots)$ with $m$ the total number of digits 0 and 1, is the size of the shortest computer programme describing it \cite{1,2,3}. The quantity $K$ is often stated to be an obviously important measure of algorithmic information. Nevertheless, it has the major disadvantage, that a purely random sequence cannot be compressed at all (it is important to note that a sequence may only be considered `random' if it is also `long enough' \cite{2}; the number of 0 digits essentially equals the number of 1 digits).

Here, I present a numerical method based on polymer physics, to get rid of the annoying random fluctuations. What purportedly remains are the long-range correlations in the data sequence that may be considered relevant variables.

First, the average $\bar{S}_m \equiv \frac{1}{m} \, \sum_{i=1}^m S_{im}$ where the $S_{im}$ denote the specific numbers of the sequence $S_m$, is subtracted from $S_{im}$, e.g. $\Delta S_{im} \!=\! S_{im} - \bar{S}_m$. The Kolmogorov complexity of the sequence of numbers (no longer the digits 0 and 1) remains the same, namely equal to $K$. Next, I introduce a quenched interaction given by 
\begin{eqnarray}
H        &=& \sum_{i \neq j}^m H_{imjm} \,, \\
H_{imjm} &=& \alpha \, \Delta S_{im} \, \Delta S_{jm} \,.
\end{eqnarray}
The strength of this interaction is denoted by the coupling parameter $\alpha$ which is scaled by the thermal energy $k_{\rm B} T$ ($k_{\rm B}$ = Boltzmann's constant, $T$ = temperature).

The next step involves the introduction of a linear three-dimensional polymer chain of $m$ contiguous spherical beads. The $i$-th bead is labelled by the number $\Delta S_{im}$. The beads are given a hard interaction $V(\vec{r}_i - \vec{r}_j)$ where $\vec{r}_i$ and $\vec{r}_j$ denote the centers of the two respective beads. The total energy of the macromolecule is given by
\begin{eqnarray}
U_{\rm tot} &=& U_{\rm hs} + k_{\rm B} T \, H \,, \\
U_{\rm hs}  &=& \sum_{i \neq j}^m \; V(\vec{r}_i - \vec{r}_j) \,.
\end{eqnarray}
Thus, in a certain configuration of the chain, two or more beads may touch each other leading to the interaction given by Eq.(3). The interaction $H$ (Eq.(1)) is also switched on upon touching of the beads. In the adhesive sphere model, Miller and Frenkel prevent divergences by introducing an effective density of states \cite{4}. They proceed by canonical simulations as first formulated by Seaton and Glandt \cite{5}. Such a scheme could be useful in our case also.

Now, if the random interactions were annealed, the following argumentation on averaging over a canonical ensemble with temperature $T$ would be straightforward. Boltzmann weighting would lead to attraction
\begin{equation}
\left< \, H_{imjm} \, \exp( - U_{\rm tot} / k_{\rm B} T) \, \right> < 0 \,,
\end{equation}
after Taylor expanding $H$ as is well known.  See for instance, the functional formalism presented in \cite{6}. The nonlinear hierarchy may be broken by a Hartree approximation leading to mean-field theories \cite{7}. In the quenched case, a general statement regarding attraction is unknown to the author even though this seems extremely plausible in general (For preliminary work, see \cite{8}). Accordingly, effective attraction is here simply hypothesized. If the coupling parameter $\alpha$ is large enough, the chain will simply collapse onto itself to form a globule of volume $V \!=\! L b^2$ where $L \!=\! m b$ is the contour length of the chain and $b$ is an effective minimum size of a bead. If there were no long-range correlations, it would be fair to assume the globule is spherical in view of isotropy. Nevertheless, there are minor fluctuations from the spherical state. 

This inevitably leads to the conclusion that the long-range correlations should give rise to deviations of the globular shape from perfect sphericity. The potato-like deviations are suggested to be correlated with the relevant information incorporated in the original sequence. The standard theory of the moment expansion of the gravitational field of the earth may then be applied \cite{9}.

In conclusion. I have conjectured a numerical way of expressing the long-range correlations within an informational sequence with the potato-like shape resulting after numerical simulations of a polymer model with a suitably chosen Hamiltonian. It would be interesting to see if any general theory may emerge from these preliminary speculations.

\vskip15pt
\noindent
{\bf Acknowledgement}
\vskip5pt
\noindent
I thank Daan Frenkel, Seth Lloyd, Martin Kr\"{o}ger, Helmut Schiessel and Kevin Dorfman for comments.

\end{document}